\documentclass[letterpaper, 10 pt, conference]{ieeeconf}  % Comment this line out
\IEEEoverridecommandlockouts                              % This command is only
\usepackage[utf8]{inputenc}
\usepackage[T1]{fontenc}
\usepackage{graphicx}
\title{\LARGE \bf
Program Controls Effectiveness Measurement Framework \& Metrics
}

\author{Abhinav Palia$^{1)}$, Caroline Devlin$^{2)}$, Megan Yelorda$^{3)}$ \\
$^{1)}$ palia@usc.edu, $^{2)}$ cdevlin6@gatech.edu, $^{3)}$ yelorda@gmail.com
}

\begin{document}

\maketitle
\thispagestyle{empty}
\pagestyle{empty}

%%%%%%%%%%%%%%%%%%%%%%%%%%%%%%%%%%%%%%%%%%%%%%%%%%%%%%%%%%%%%%%%%%%%%%%%%%%%%%%%
\begin{abstract}

Any program that is designed to accomplish certain objectives, needs to establish program level controls pertaining to the overall goal. A critical aspect that determines the success of a program is the quality of the controls and their effectiveness in accomplishing the goal. Traditional Control Maturity Models primarily focus on the efficiency, management, and optimization of controls, while only indirectly measuring control effectiveness which neglects an essential aspect of control efficacy. In this paper, we highlight the ineffectiveness of these models, outline an adaptable program controls framework, and provide an approach to define measurable attributes (metrics) that enable a zero-defect program. To the best of our knowledge, we believe this is the first paper that provides a structured approach to defining a controls measurement framework and creation of effectiveness metrics that can be adopted by a variety of program use cases.

\end{abstract}

%%%%%%%%%%%%%%%%%%%%%%%%%%%%%%%%%%%%%%%%%%%%%%%%%%%%%%%%%%%%%%%%%%%%%%%%%%%%%%%%
\section{INTRODUCTION}

The operationalization of a zero-defect program involves establishing program goals and objectives, identifying the right activities to accomplish them, and having checks in place to ensure implementation. Over the period of time, the management teams responsible for a program’s development and maintenance, focus primarily on the maturity of controls and generally, try to follow the traditional maturity models focusing on the efficiency, governance and optimization of controls. However, one of the most critical aspects of measuring program controls effectiveness is often overlooked. Determining control effectiveness is paramount to defining the qualification of controls in the program and as a precursor to engaging in maturity efforts. It is crucial to analyze the quality of the controls defined and how effectively they support the ability to achieve the overall program goals. Several frameworks and guidance~\cite{c1},~\cite{c4}-\cite{c6}, proposed to measure effectiveness of the program controls are either specifically designed for technical programs, limiting their applicability to non-technical programs or are subjective and non-quantifiable, making them impractical for many program use cases. 
\par In this paper, we have defined a generic program controls framework focusing on measuring controls effectiveness. This framework is designed to be flexible so as to fit a variety of program use cases. Using this framework, we provide an approach and guidance on how to develop quantifiable metrics to measure the quality of controls which provides the basis for exclusion or inclusion of controls in the program. Once the program controls have been selected and tested against the metrics, the next step is to define a program controls maturity matrix in order to make the controls efficient and cost effective. After the program controls have attained the highest maturity, program management and leadership should strive for controls optimization. Combining all these concepts, we can define the overall program success as a function of optimized and effective controls. We understand that even after the execution of these effective and optimized controls, there are certain unidentified risks that could exist and pose danger to the program goal accomplishment but we consider them to be beyond the scope of this paper. 
\par The rest of the paper is organized as follows - in the next section, we review the related work in the area of program controls maturity and effectiveness. In section III, we describe an overview of the program controls effectiveness measurement framework followed by guidance to define metrics for the measurement of controls effectiveness in section IV. Then in section V, we present a discussion to define overall program success and in section VI, we provide some unique examples demonstrating the application of our framework and finally, summarize and conclude this paper.

\section{Related Work}

To establish a successful program, it is critical that the program framework provides flexibility in its approach to support the structure and scope of the program, and dynamic in its ability to utilize controls effectiveness metrics to improve upon itself. Prior to the implementation of this program controls effectiveness and measurement framework, we investigated and reviewed commonly used technical frameworks relative to our research to determine their applicability to several unique use cases.
The National Institute of Standards and Technology (NIST) Cybersecurity Framework~\cite{c1} describes five general categorizations for program level controls; identification, protection, detection, response, and recovery. Within each category, the framework describes sub-categories which define specific activities targeted at accomplishing the goal of the overarching function. While this framework outlines valid categorizations for the minimization of risk within cybersecurity and provides details for the implementation of technical safeguards, the framework fails to provide direction for measuring the effectiveness of technical safeguards for mitigating risk. Further, it fails to identify an approach for the prioritization of program level controls, creating a technical structure that is both too rigid and too broad to provide a clear method for guaranteeing a program’s general success.

\par The Capability Maturity Model Integration (CMMI) Framework~\cite{c2} aims to provide value for program performance in terms of increased efficiency and effectiveness, quality, and communication following a maturity scale defined by specific, program level activities. Although this framework provides a general structure for how to categorize the maturity of program level controls, it does not identify the distinction between the effectiveness of specific controls and the maturity of the program as a whole. The CMMI Framework also does not identify how controls are tracked and measured to drive this maturity and determine thresholds of control effectiveness unique to a specific program, forgoing the fundamental operationalization required for determining the success of a program.
\par \cite{c3} outlines a set of ten factors that determine the level of adequacy with which a set of program level controls can address risk. This framework sacrifices effectiveness for efficiency, stressing the importance of control completion within a program over control operationalization and quality within a program. \cite{c3}, therefore, provides guidance solely surrounding the implementation of controls and lacks adequate structure for the measurement of effectiveness of controls.
Ernst \& Young’s approach for building confidence in internal program level controls~\cite{c4} assesses a private company’s initial need for improved control effectiveness by determining the current maturity of a program’s existing framework, and then compares the identified risks associated with particular program controls against their inventory of a program’s industry best-practices. In outlining a program’s structure in this way, it creates a level of program dependency as a program’s success is measured according to their singular standard of effectiveness. This rigidity and lack of program personalization implies that a program cannot be mature nor effective if the program level controls, and risks associated with such controls cannot be mapped to E\&Y’s internal suite of key performance indicators. 
\par The Outline for the Assessment of Control Effectiveness presented in~\cite{c5}, provides a series of questions aimed at thinking critically about the priorities of a program, specifically tenets surrounding internal staff evaluation, objectives and performance indicators, and procedure documentation and management. The approach outlined in [5] lacks the capability to define quantifiable metrics for measuring control effectiveness, solely identifying whether certain controls or technical gaps exist within a program. 

\par In~\cite{c6}, three primary use cases for the measurement of control effectiveness and two approaches for qualitatively measuring this effectiveness are defined. It provides loose definitions for these approaches, supplying inadequate evidence or data to support the claim that the approaches can truly quantify effectiveness. It leaves a large portion of the framework to the discretion of the program creators and fails to define thresholds for maturity or continued program improvement.
Next, we define a dynamic approach to creating a unique program controls framework, and the effectiveness metrics associated with such controls.

\section{Overview of Program Control Effectiveness Measurement Framework}

In order to have a successful program, it is critical to identify a  primary goal as well as secondary, subsidiary goals. The program should define tenets to provide guiding principles required in accomplishing program goals. These principles should be able to support, lay the boundaries and scope of the program. Next, activities should be listed to attain these goals. These activities essentially are the program level controls which serve to operationalize the realization of the program objective. The program level controls should be tracked for completion and maturity levels should be defined. This enables the guidance for continuous improvement of controls efficiency. The most critical aspect of a program quality measurement is to determine if the controls defined are effective in attaining the primary goal of the program. Defining and calculating the metrics to measure the quality of the operationalization efforts - the program level controls, helps understand the discrepancies in the program and improve upon them. The management and maintenance of the program spans over all of the controls and is responsible for reviewing the program regularly to identify and analyze the trends in the metrics, perform the root cause analysis, suggest and implement mitigation strategies. 
\par The program controls maturity along with the controls effectiveness determine optimum success of the overall program controls. The degree of effectiveness of the controls can be bound by program specific cutoffs for the metrics in congruence with the maturity level as discussed in section V. Over the period of time, maturity stabilizes and reflects optimality of the controls. However, the effectiveness can increase or decrease in reference to the cutoffs, reflecting the need to alter the controls or define supplementary controls to increase effectiveness and hence, achieve program objectives.  
\begin{figure}[t]
   \centerline{\includegraphics[width=70mm, scale=0.5]{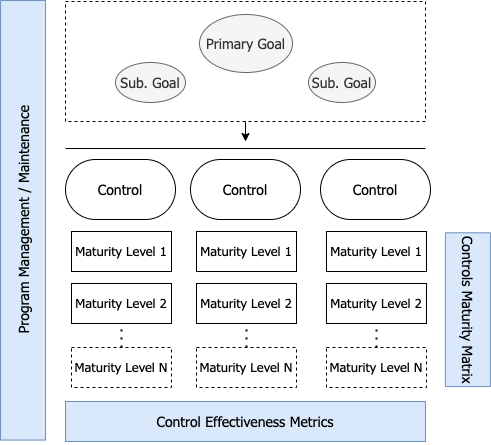}}
    \caption{An Overview of Controls Effectiveness Measurement Framework}
    \label{fig:cefoverview}
\end{figure}

\par Fig.\ref{fig:cefoverview} represents a layout of a program having primary and subsidiary goals, which form the basis for the controls required to accomplish the objective of the program with multiple levels of maturity represented by a maturity matrix. Controls Effectiveness spans over all the controls and the management/ maintenance layer oversees the entire program. In the next section, we discuss the approach to define the quantifiable metrics for measuring the quality and effectiveness of the program level controls. 

\section{Metrics to Measure Effectiveness}
\begin{figure}[t]
  \centerline{\includegraphics[width=70mm]{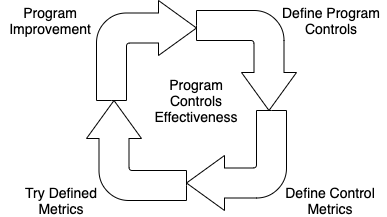}}
    \caption{Program Controls Effectiveness Metrics driving Program improvement}
    \label{fig:flywheel}
\end{figure}

Defining meaningful, quantifiable metrics that represent a true measure of quality of the controls established for a program is both challenging and one of the most critical aspects measuring the effectiveness of a  program. In order to do this, the primary goal and the subsidiary goals of the program must be the main focal points along with the program tenets to outline the scope. It is critical that the defined metrics not only measure the current state of the program activities, but should have the capability to envision and drive future program improvement. 
Prior to the identification and definition of effectiveness metrics, all program controls aimed at accomplishing the program goals must be thoroughly reviewed, understood, and agreed upon by program managers. Once the program controls have been established, research should be done to determine if there are any commonly held quantity or quality metrics that should be applied based on their congruency and applicability to the program.  For each control, the underlying outcome should be identified and all the possible quantifiable measures driving the success and failure of the outcome should be listed. Using a trial and error approach, this exercise should produce an exhaustive list of metrics. Next, it should be filtered by considering the following four benchmark questions for each metric: 
\begin{enumerate}
    \item How does the metric measure the quality of a particular program control or program control categorization?
    \item How does the metric relate to the primary goal, subsidiary goal, or a program tenet?
    \item How will the data for measuring the metric be obtained, calculated, and presented?
    \item How will the metric can be used to drive further program improvement?
\end{enumerate}

If a proposed metric does not meet the criteria outlined by the four benchmark questions, it indicates that the current state for that metric provides little to no insight into the quality of a particular program control. It must therefore be re-framed or dismissed from consideration. In the event that a thorough definition can be provided for each of these four benchmark factors we can conclude that the metric represents the effectiveness of a program control and should be qualified for initial implementation. The trial phase of implementation serves as the observation period used to generate data points as a step in testing the qualitative, quantifiable nature of a particular effectiveness metric. Testing determines whether a metric is useful and measurable in practice. Upon observation assuming that the metrics that provide insights on effective operationalization of program controls or highlight gaps for program improvement (as represented in Figure \ref{fig:flywheel}), they are finalized as the Control Effectiveness metrics for the program. 
\par In the next section, we discuss the technical formulation for a program, corresponding controls effectiveness metrics, and a maturity matrix, which together support how to define the overall scope of success and failure of a program. 

\section{Discussion}
For any Program $P$, defined by a controls framework $C = \{ C_1, C_2,\ldots\, C_p \}$, where $C_i$ represents the controls to achieve the goal $G$ of the program. Controls Effectiveness metrics for $P$ that qualify for the observation period (as discussed in the previous section) can be defined as $M = \{M_1, M_2,\ldots\ , M_q\}$, and a maturity matrix as $X$, together can help define the overall scope of success and failure of $C$. The values for $M(C_i)$ represent the degree to which a control is effective and can be bound by a lower threshold $t$ as defined by $P$, to identify what controls are not effective given the right context and application as a function of time duration $\tau$. However, for every $M_k$, the measurement scale is application specific and is bound by an upper threshold $T$ as the management tries to minimize or maximize the metric values for the controls. In summary, the scale $[t, T]$ is used to define the degree of effectiveness. It is common to use fuzzy values for $[t, T]$, such as \textit{partially effective}, \textit{effective}, and \textit{extremely effective} to represent the effectiveness instead of using numeric values. The maturity of controls is dependent on the nature of the control and after a certain time duration $\tau$', and attaining the $max. \{X(C)\}$ then $X \to O$, where $O$ represents the optimization of program controls. 
\par In conclusion, effectiveness as a function of metrics $M$, helps in determining the qualification of a control $C_i$, in achieving $G$, and the maturity matrix $X$ for $C$, as a function of time, focuses on the efficiency. After attaining maximum maturity, $X$ reaches for optimization, as against the effectiveness which bounds the quality of controls in achieving the goal making it more crucial for a program. Ideally, determining controls effectiveness should be done prior engaging in maturity efforts, as the former provides a strong basis and proof of concept for including a control or adding supplementary controls in the program. 
\begin{figure}[t]
  \centerline{\includegraphics[width=70mm]{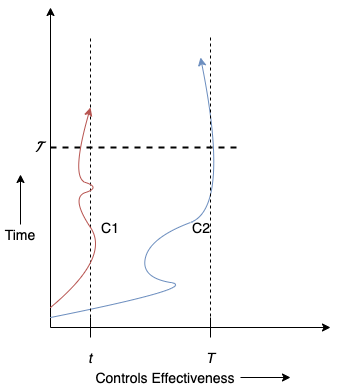}}
    \caption{Program Controls Effectiveness Metrics driving Program improvement}
    \label{fig:controls-discussion}
\end{figure}
\par Figure \ref{fig:controls-discussion} represents two controls $C_1$, and $C_2$, which are being observed for effectiveness using the metrics defined on the scale of $[t, T]$ for the trial phase (observation time $\tau$). Evidently, $C_2$ is helping achieve the program goal while $C_1$ is not, therefore guiding the management to restrict resources being spent in $C_1$ and eventually exclude it from the program. 
\par For determining the scale for the metric values per control, management usually strives for maximizing or minimizing the metric values, and gathering enough data points during the trial phase (for the active controls). Note, this observation period is dependent on the nature of the program and in some cases, metric observation can be an ongoing process as long as it is generating relevant data. However, the threshold values can be tuned in order to provide a buffer for controls inclusion. In order to account for anomalies, the management layer has to analyze the trends over this observation period before deciding the fate of a control.
\par The overall success $\psi$ of a program $P$, can be achieved when optimum resources are spent to implement controls $C$ in order to achieve the desired goal $G$ of the program. In other words,\

\begin{center}
Given,$\ \bigcap C=\emptyset \ and \ \bigcup [\rho(C)] = G$, then 
\\
$\psi(P)$ = $f(O(C)) \mid \forall C, M(C) > t$ 
\end{center}

where, $O$ is the optimization function which is applied on all the effective controls, given that all controls in $C$ are mutually exclusive and the union of implementation outcome ($\rho$) of the controls achieves $G$. Function $f$ accounts for the unknowns.

\section{Example Use Cases}
In this section, we will present an application of our program controls effectiveness framework in the following unique use cases and discuss the relevance of the defined approach. 
\begin{itemize}

\item {\bf HIPAA Compliance Program}: In order to enable a zero-defect policy for a compliance program and protect unsecured access to PHI systems by third party consultants while building customer systems, this program is designed with controls such as educating the customers and consultants, performing periodic risk evaluations, providing compliance guidance while building architectures, on-boarding and off-boarding. The primary program goal is to reduce the overall risk, and the efficacy of metrics is defined using metrics such as knowledge comprehension over time, frequency of a particular finding as an outcome of risk evaluations, and number of suspected incidents, mapping back to the controls and their impact in reducing the HIPAA violation risk. During the observation period, the controls will be refined to generate meaningful data points and decisions on the inclusion / exclusion of controls are made. Once the initial controls and metrics are selected, the next step is to make them efficient and optimize controls (for example, automating the manual tasks of assessment, having online study material for education, etc.). 
\item {\bf Fashion Marketing Program}: The primary goal of a fashion marketing program is to maximize sales and the subsidiary goal is brand awareness. The controls defined to achieve these goals include advertisement on several platforms, email marketing, photo shoot endorsement, etc. Defining metrics for a program like this could be challenging if a proper methodology is not followed. However, using the approach defined in this paper, the metrics can be created to quantitatively measure the quality of these controls such as, website visit via email clicks, sales driven by photo shoot endorsements, and sales driven by marketing efforts. In terms of maturity of the controls, automation of analytics and machine learning can be used to assist decision making therefore leading to a successful program.

\end{itemize}

\section{CONCLUSION}

In this paper we demonstrate that measuring control effectiveness is a gating factor to achieving program goals and objectives. Traditional control frameworks mostly fall short in providing a structured approach to define quantifiable metrics for measuring the quality of program controls. Also, most of the frameworks only address the maturity of program controls and lack a flexibility to fit a varied set of use cases. We presented a general program controls framework and provided guidance on deriving and creating metrics for program controls effectiveness. We described overall program success as a function of controls effectiveness and maturity, and presented the applicability of our framework in two unique use cases. To the best of our knowledge, this is the first paper elucidating a practical strategy for producing measurable metrics that provide insights on effective operationalization of program controls and highlight gaps for program improvement.

\addtolength{\textheight}{-12cm}   % This command serves to balance the column lengths
                                  % on the last page of the document manually. It shortens
                                  % the textheight of the last page by a suitable amount.
                                  % This command does not take effect until the next page
                                  % so it should come on the page before the last. Make
                                  % sure that you do not shorten the textheight too much.

%%%%%%%%%%%%%%%%%%%%%%%%%%%%%%%%%%%%%%%%%%%%%%%%%%%%%%%%%%%%%%%%%%%%%%%%%%%%%%%%

%%%%%%%%%%%%%%%%%%%%%%%%%%%%%%%%%%%%%%%%%%%%%%%%%%%%%%%%%%%%%%%%%%%%%%%%%%%%%%%%

%%%%%%%%%%%%%%%%%%%%%%%%%%%%%%%%%%%%%%%%%%%%%%%%%%%%%%%%%%%%%%%%%%%%%%%%%%%%%%%%

%%%%%%%%%%%%%%%%%%%%%%%%%%%%%%%%%%%%%%%%%%%%%%%%%%%%%%%%%%%%%%%%%%%%%%%%%%%%%%%%

\end{document}